\documentstyle[overcite,prl,aps]{revtex}
\input epsf

\def\ba{\begin{eqnarray}}
\def\ea{\end{eqnarray}}
\def\be{\begin{equation}}
\def\ee{\end{equation}}

\newcommand{\labeq}[1] {\label{eq:#1}}

\def\mxth{\mathsurround=0pt }
\def\xversim#1#2{\lower2.pt\vbox{\baselineskip0pt \lineskip-.2pt
    \ialign{$\mxth#1\hfil##\hfil$\crcr#2\crcr\sim\crcr}}}

		    \def\gsim{\mathrel{\mathpalette\xversim >}}

\begin{document}

\twocolumn[\hsize\textwidth\columnwidth\hsize\csname@twocolumnfalse\endcsname

\title{Is Vacuum Decay Significant in  Ekpyrotic and Cyclic Models?
 }

\author{Paul J. Steinhardt$^1$ and Neil Turok $^{2}$}

\address{
$^1$ Joseph Henry Laboratories,
Princeton University,
Princeton, NJ 08544, USA \\
$^2$ DAMTP, CMS, Wilberforce Road, Cambridge, CB3 0WA, UK}

\maketitle

\begin{abstract}
It has recently been argued that bubble
nucleation in ekpyrotic and cyclic cosmological scenarios can
lead to unacceptable inhomogeneities unless certain constraints are
satisfied.  In this paper we show that this is not the case.  We find
 that bubble nucleation is completely negligible in realistic models.
\end{abstract}
]

The cyclic model of the Universe\cite{ST1,ST2} 
is a radical alternative to standard big bang and inflationary 
cosmology. The cyclic picture
 proposes that the Universe undergoes an 
endless sequence of cosmic epochs which begin with expansion
from a big bang and end in collapse to a big crunch.
The connection between the big crunch and the ensuing 
big bang is presumed to occur according to the prescription
recently proposed with Khoury, Ovrut, Seiberg.\cite{nonsing}
The expansion part of each cycle has a period of radiation 
and matter domination followed by an extended period of 
cosmic acceleration at low energies.  The long period of
acceleration is crucial in establishing the flat and vacuous 
initial conditions required prior to beginning the contraction
phase.  The density of 
entropy, black holes and other debris from the previous 
cycle is reduced to nearly zero.  
Subsequently, the Universe ceases to accelerate, fluctuations 
are generated, and the Universe heads towards a big crunch
in which matter and energy are regenerated. 
The key ingredient is the restoration of
the Universe
to a pristine flat vacuum state before each big crunch,  
which  ensures that the cycle can repeat and that the cyclic 
solution is an attractor.

The cyclic model can be described in terms of a scalar field
$\phi$
moving back and forth in an effective potential $V(\phi)$ 
(see Fig. 1). Although the model can be described in terms of
an ordinary quantum field in four dimensions,
the model is strongly motivated by ideas developed as part of 
the ekpyrotic scenario,\cite{kost1}  
drawing on the braneworld picture and string theory.
The scalar field represents the modulus field that determines
the distance between branes. Moving from some positive
value to minus infinity and back along the potential
corresponds to a pair of orbifold branes being drawn 
together by an interbrane force, colliding (corresponding 
to $\phi \rightarrow - \infty$), bouncing and 
rebounding.  The transition between big crunch and big 
bang corresponds to the bounce.\cite{nonsing}
The accelerating phase  begins when the branes are maximally 
separated ($\phi=\phi_0$)
once the matter density falls below the positive
interbrane potential energy density.
 As the branes are drawn together,
the acceleration rate decreases until, at $\phi= \phi_1$, the 
expansion begins to decelerate. As the branes draw yet 
further together,  the potential becomes negative ($\phi<0$), the 
expansion stops ($\phi=\phi_2$) and the universe begins 
to contract.  Fluctuations are generated during the 
contraction phase ($\phi=\phi_3$), before the potential 
reaches a minimum at $\phi=\phi_{min}$.

\begin{figure}
\begin{center}
 \epsfxsize=3.5 in \centerline{\epsfbox{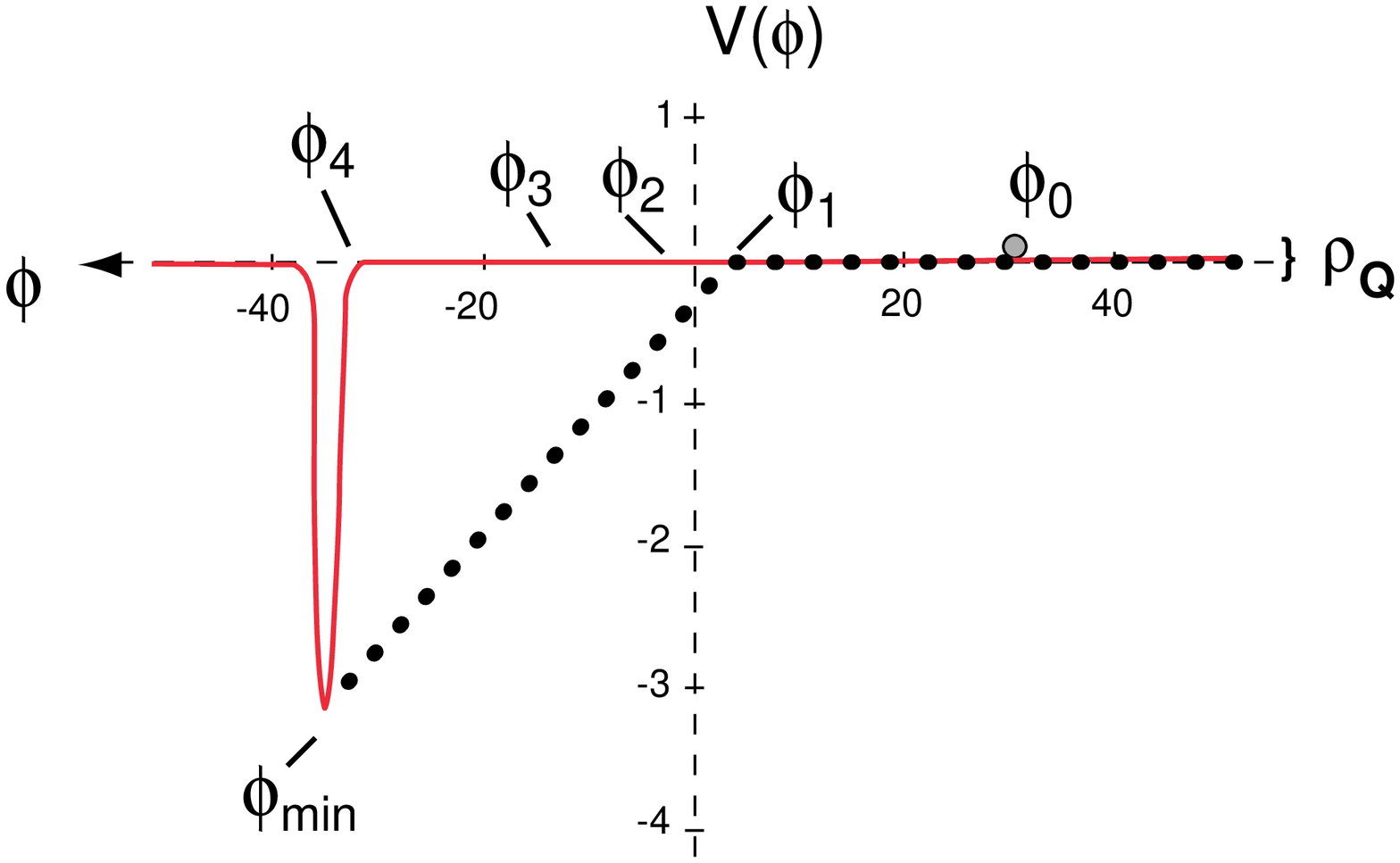}}
  \end{center}
    \caption{
     Plot of the actual analytic  potential (solid line)
     described in Refs.~1 and 2, shown to proper scale.
     For the purposes of analyzing bubble nucleation,
     differences between the schematic plots shown in
     Refs.~1 and~2 and the properly scaled plot are
     significant.  The axes are expressed
    in Planck  units. The value of $\rho_Q$ is about $+10^{-120}$
    corresponding to the current dark energy (quintessence)
    energy density.  The markings $\phi_i$  refer to 
    points described in the text.  The dotted line is 
    the piecewise linear approximation assumed by 
    Heyl and Loeb. Our central point is that the piece-wise
    linear function
    is an exponentially
    poor approximation to the slope
    for $\phi < \phi_1$, leading to an exponential overestimate
    of the nucleation rate.  If the kink is moved to $\phi_4$,
    the piecewise approximation is reasonable, but the decay rate is 
    then negligible.
		 }
		    \end{figure}

Recently, Heyl and Loeb\cite{HL} have argued that bubble nucleation 
during the  accelerating and collapse phase, near $\phi= 
\phi_1$, for example, can spoil the 
homogeneity of the universe unless the potential obeys 
certain constraints.  They have focused on tunneling
about $\phi_1$, where the 
the kinetic and potential energy become comparable as the field
rolls to the left and where the accelerating expansion phase ends.
Bubbles form whose interior corresponds
to values of $\phi$ far down the potential.  Once formed,
the bubbles grow at the speed of light.)
These bubbles destroy the homogeneity essential to the scenario.
Their  point is that, although the bubbles are rare events,
there remain 
more than a billion years from when $\phi$ passes $\phi_1$
until
it reaches the bottom of the potential well. They estimate
that this is sufficient time to have a significant amount of 
bubble nucleation, unless  the potential well is not very deep.

In this paper we show that the approximations used by Heyl 
 and Loeb,  although  motivated by schematic figures of
 the potentials in Refs.~\ref{ST1} and ~\ref{ST2},
 are not valid for the actual analytic forms, as
 presented  for  either ekpyrotic or cyclic models.
In particular, they estimate the bubble nucleation rate
assuming a potential which can be approximated as two 
linear segments, one with slope zero and one with non-zero
slope  (Fig. 1).  
However, the slope of the potentials used in the ekpyrotic
and cyclic models is $10^{100}$ or so times smaller
than in their piece-wise linear approximation 
 for $\phi_4 \ll \phi < \phi_1$. 
This error is further amplified in the computation of the 
bubble nucleation rate, which is exponentially sensitive to 
the slope.  The smaller is the slope, the more suppressed is
the nucleation rate.  For the potentials actually used, we will show that
the bubble
nucleation rate is completely negligible compared to the classical
rolling rate.

The  explicit model for $V(\phi)$ which was discussed in
the context of the cyclic model is of the form:\cite{ST1,ST2}
\be
V(\phi)= V_0(1 - e^{-c \phi}) F(\phi),
\labeq{examplep}
\ee
where, without loss of generality,
we have shifted $\phi$ so that the zero of the potential
occurs at $\phi=0$.
The function $F(\phi)$  is introduced to
represent the vanishing of non-perturbative effects
described above: $F(\phi)$
turns off the potential rapidly as $\phi$ goes below
$\phi_{min}$, but it approaches one for $\phi > \phi_{min}$.
For example, $F(\phi)$  might be proportional
to $e^{-1/g_s^2} $   or $e^{-1/g_s}$ (for illustration, we will
assume the latter form),
where $g_s \propto e^{\gamma \phi}$ for $\gamma > 0$.
We will choose $c = 10$ and $\gamma \approx 1/8$ so that the potential
minimum corresponds to $V(\phi_{min}) = M_0^4 = {\cal O}(1)$ in
Planck units, a choice which Heyl and Loeb claim is ruled out  by the
nucleation constraint.
The constant $V_0$ is set roughly equal to
the vacuum energy observed in today's Universe, of order
$10^{-120}$ in Planck units. 
 For large $c$,
this potential has $V''/V \ll 1$ for $\phi \gsim 1$ and $V''/V \gg 1$
for $\phi_{min} < \phi<0$. 
 These two regions  account
for cosmic acceleration and for ekpyrotic production of
density perturbations,  respectively.\cite{kost1,ekperts}
More precisely, the end of acceleration is at $\phi_1  \approx
0.35$. The
 rest of the markers in Figure~1 are scaled to the same parameters.
 The current value (circle) corresponds to tens of trillions of 
 years before the big crunch.
In the ekpyrotic  region, the constant term is irrelevant and
$V$ may be approximated by
$ -V_0 \, e^{-c \phi}$.  
The scaling solution over this range is
\be
a(t)= |t|^p,
\qquad V=-V_0e^{-c \phi} = -{p(1-3p)\over t^2},
\qquad
p={2\over c^2}.
\labeq{back}
\ee
Here we chosen  $t=0$ to be the bounce.  These expressions are
 used to obtain the time estimates discussed in the paper (and 
 this agrees with Heyl and Loeb).

Heyl and Loeb focus on $\phi$ near $\phi_1$, where the 
scaling solution first becomes valid. The Universe is expanding
but at a decelerating rate. As the field roll towards more negative 
values, the potential energy  becomes increasingly negative,
the expansion stops  and contraction begins.
They approximate the potential as 
\begin{equation}
V(\phi) = - M_0^4 \frac{\phi- \phi_1}{\phi_{min}- \phi_1} 
\Theta(\phi_1 -\phi)
\end{equation}
where $\Theta(\phi)$ is the Heaviside step function.
This amounts to approximating $V(\phi)=0$ for $\phi > \phi_1$
and $V(\phi) =  K \cdot (\phi- \phi_1)$ for $\phi< \phi_1$, where
$K$ is a constant 
chosen to be the slope of the line  that connects $\phi_1$ to
$\phi_{min}$. We have superposed this approximate form in
Figure 1.    The piecewise linear potential was introduced
by Lee and Weinberg in their analysis of bubble nucleation 
for a rolling scalar field without a barrier.\cite{LW}

There are two important aspects to the bubble nucleation 
problem in the piece-wise linear potential. First,
as derived by Heyl and Loeb,\cite{HL} gravitational effects suppress
the tunneling altogether unless $\phi$ is within a Planck
distance of $\phi_1$, $\eta = \phi- \phi_1 < {\cal O}(1)$.
Hence, for values of $\phi$ far to the right of $\phi_1$ 
on the plateau, tunneling past $\phi_1$ is insignificant even though the
classical rolling rate is slow.
 Second, as derived by Lee and 
Weinberg, the bubble interior corresponds to a jump in $\phi$
in which $\phi$ jumps past $\phi_1$ by an amount 
$\eta$. So, the closer is $\phi$ to $\phi_1$, the less
is the jump.  
Combining the two aspects, it is clear that  a 
good 
piece-wise linear approximant to $V(\phi)$ to be used to estimate
tunneling near $\phi_1$
 should match the slope near $\phi_1$
within one Planck distance, $\Delta \phi \le 1$.
The behavior of $V(\phi)$ for 
values  farther away in either direction plays no role 
in the nucleation rate about $\phi_1$.

As is evident from Fig.~1, 
 the slope of the piecewise 
linear approximation  disagrees by an exponential factor from 
the actual slope of the potential for $\phi < \phi_1$.
Within a Planck distance of $\phi_1$,
the difference in the actual slope and the piecewise approximant
is a factor of nearly $10^{100}$. The difference totally changes
the conclusion.  The bubble nucleation rate
is proportional to\cite{LW}
\begin{equation}
{\rm exp} \, \left[ - \frac{32 \pi^2}{3} \frac{(\phi- \phi_1)^3}{K}
\right].
\end{equation}
That is, the exponentially smaller slope of the actual potential 
is further exponentiated in computing the nucleation rate.
Consequently, the nucleation rate is exponentially smaller than
the estimate obtained by Heyl and Loeb and is completely negligible,
as we suggested in our paper.\cite{ST2} 

The argument is not complete.
We have argued  that the piecewise linear 
approximation is poor near $\phi=\phi_1$.  However, one might
worry that it  is a  reasonable  approximation
for $\phi$ near $\phi_4$,
where the potential becomes steeper.
Does nucleation become significant  as $\phi$ approaches $\phi_4$?
First, note that  Heyl and Loeb's result can be generalized to
show that
nucleation is insignificant 
if
\begin{equation}
(\phi - \phi_4)^3 (\phi_4 - \phi_{min}) \gg M_0^4
\end{equation}
where $-M_0^4$ corresponds to the minimum of the potential.  
Here we have chosen $M_0$ in Planck  units.  In this case,
where $\phi_4  - \phi_{min} \approx 3$, we can ignore 
nucleation unless $\phi$ is within a few Planck distances of $\phi_4$.
The stronger condition, cited above, is that gravitational effects
suppress nucleation unless $\phi$ is less than one Planck unit
from $\phi_4$.  
Then, the nucleation rate is comparable to what Heyl and Loeb 
found.  However, the time left before the field rolls to 
potential minimum is exponentially smaller.
At $\phi \approx \phi_1$, the remaining time is about
$H_0^{-1} = 10^{60}$ in Planck units.
For $\phi$ within less than a Planck distance from
$\phi_4$, the remaining time is less than 
one Planck time. There is not enough time left to have 
significant chance to nucleate because here the field is rolling
too quickly.

If one considers tunneling near points between $\phi_1$ and
$\phi_0$ (today's value),
the tunneling rate is more suppressed
than at $\phi_1$ because the slope of the potential $K$ is less.
The rolling time increases as $(V'')^{-1} \approx  1/c K$ but 
the tunneling time increases by $\sim {\rm exp}(1/K)$, 
an exponentially greater factor.

We confess to bearing  some responsibility for Heyl and Loeb's 
mis-application of the piecewise linear approximation to the potential.
We provided in our papers\cite{ST1,ST2} a schematic drawing 
of the potential which suggests that the slope near $\phi=\phi_1$ 
(where there is a billion years before reaching the minimum)
is steeper than is actually the case.
This is apparent if one plots the actual potential function described
in the text, as has been done in Figure 1.

The bottom line is that bubble nucleation is negligible throughout the
cyclic solution and provides
no significant contraint for the potentials we consider in ekpyrotic
and cyclic models.  However, it is good to be aware of Heyl and 
Loeb's analysis if one considers more general potentials.

We thank J. Khoury, J. Heyl and A. Loeb for useful discussion and
for reviewing the manuscript. 
This work was supported in part by
 US Department of Energy grant
DE-FG02-91ER40671 (PJS) and by PPARC-UK (NT).

\end{document}